\documentclass[prl,twocolumn,floatfix,showpacs,showkeys]{revtex4} 
\usepackage[first,draft]{draftcopy}
\usepackage{latexsym,amssymb,amsmath,amsfonts}
\usepackage{graphicx}
\usepackage{bm}
\newcommand{\beq}{\begin{equation}}
\newcommand{\eeq}{\end{equation}}
\newcommand{\bc}{\begin{center}}
\newcommand{\ec}{\end{center}}
\newcommand{\eeqa}{\end{eqnarray}}
\newcommand{\beqa}{\begin{eqnarray}}
\newcommand{\no}{\noindent}
\newcommand{\pa}{\partial}

\newcommand{\ra}{\rightarrow}
\newcommand{\na}{\nabla}

\newcommand{\al}{\alpha}
\newcommand{\be}{\beta}
\newcommand{\ga}{\gamma}
\newcommand{\Ga}{\Gamma}
\newcommand{\de}{\delta}

\newcommand{\ep}{\epsilon}

\newcommand{\et}{\eta}
\newcommand{\ka}{\kappa}
\newcommand{\la}{\lambda}

\newcommand{\rh}{\rho}
\newcommand{\si}{\sigma}
\newcommand{\Si}{\Sigma}

\newcommand{\ph}{\phi}

\newcommand{\ps}{\psi}

\newcommand{\om}{\omega}
\newcommand{\ed}{\end{document} }

\begin{document}

\title{Spin flip probability of electron due to torsional wave}
\author{Richard T. Hammond}

\email{rhammond@email.unc.edu }
\affiliation{Department of Physics\\
University of North Carolina at Chapel Hill\\
Chapel Hill, North Carolina and\\
Army Research Office\\
Research Triangle Park, North Carolina}

\date{\today}


\begin{abstract}
The probability of spin flip of an electron due to a torsional wave is calculated. It is compared to the electromagnetic case, and ways to detect torsion are discussed.
\end{abstract}

\maketitle
   
\section{Introduction}

In modern formulations of gravity, from supergravity and string theory to local gauge formulations, torsion is a necessary ingredient, for reviews of torsion see the literature.\cite{hehl},\cite{hammondreview},\cite{shapiro} Soon after Einstein published the general theory of relativity in 1915, the limited geometry developed by Riemann was seen to be a small part of the full geometry of curved spacetime,  which is now known to include torsion and non-metricity (although non-metricity is assumed to vanish here). It has been shown in \cite{hammondreview} that torsion arising from the intrinsic spin of particle leads to the correct conservation law for total angular momentum plus spin, and more recently it was shown that torsion must exist on local gauge theory principles.\cite{hammondprize}

Over the years there have been many mathematical formulations of gravity with torsion, but the most generally agreed upon theories  are those where the source of torsion is the intrinsic spin of an elementary particle. Orbital angular momentum does not create torsion, and can be accounted for in the usual way with the conventional energy momentum tensor describing rotating matter. Such a situation manifests itself within the metric tensor, but does not give rise to torsion.

In the literature there are two main approaches to the spin formulation of gravity with torsion. One is the local Poincar\'e gauge theory in which the torsion is the gauge potential,\cite{hehl} and the other is the string theory-like torsion in which case the torsion is derived from a potential 2-form. In this article we assume torsion is derived from a potential.

On the experimental front, most efforts search for a small dipole field
created by spin-polarized materials acting on particles 
with intrinsic spin.\cite{ni},\cite{qiu},\cite{hammondprd} So far, the best these experiments can do is place an upper limit on the coupling constant.
However, torsion can manifest in other ways and the point of this paper is to describe an entirely new approach to detect it.

The approach is this: First a torsion wave will be created by coherently spin flipping a large number of particles with intrinsic spin. This is the source. For example, electrons in an alternating electric field will suffer such flipping. This will induce other electrons, the detector, to flip. This flip can be measured by conventional electrodynamic means. In the present paper, we calculate the probability that the torsion wave flips the electron, so we can ultimately learn how much power is needed to detect the field.

Potential future measurements of torsion will be a groundbreaking event. Not only will it present the first demonstration of non-Reimannian geometry, it may also be interpreted as the first measurement of an effect predicted by string theory. Natural units are used mostly, but occasionally cgs units are inserted for clarity.

\section{Background}

The field equations for gravity and torsion may be derived from the action principle

\beq\label{action}
\de\int d^4x\sqrt{-g}\left(\frac{R}{2k}+L\right)=0
\eeq

\no where $k=8\pi G$ and $R$ is the curvature scalar of $U^4$ space, i. e., spacetime with torsion, and $L$ represents matter coupling.

In order to get a physical field, i.e., a field obeying a second order differential equation, it is assumed that the torsion is derived from a potential $\ps_{\mu\nu}$ according to

\beq\label{torpotdef}
S_{\mu\nu\si}=\psi_{[\mu\nu,\si]}=\frac13
(\psi_{\mu\nu,\si}+\psi_{\si\mu,\nu}+\psi_{\nu\si,\mu})
.\eeq

\no This is also referred to as string theory torsion and the Kalb-Ramond field. Discussion of the history of this field, which predates string theory, may be found in the literature.\cite{hammondreview}

The field equations are

\beq\label{fe}
G^{\mu \nu}  - 3S^{\mu \nu \sigma}_{\ \ \ \ ; \sigma}
-2S^{\mu}_{\ \alpha \beta}S^{\nu \alpha \beta} = kT^{\mu \nu}
\eeq

\no where $G^{\mu \nu}$ is the (non-symmetric) Einstein tensor of $U_4$ spacetime.
The  torsional field equations, which are what we are interested here, are the antisymmetric part of this,

\beq\label{tfe}
S^{\mu \nu \sigma}_{\ \ \ \ ;\sigma} =-kj^{\mu \nu}
\eeq
where $j^{\mu\nu}\equiv (1/2)T^{[\mu\nu]}$.

It is also useful to write the equations in $V_4$ (Riemannian) spacetime

\beq\label{gr+}
^oG^{\mu\nu}=k(T^{\mu\nu}+t^{\mu\nu})
\eeq
where
\beq
kt^{\mu\nu}=3S^{\mu\al\be}S^\nu_{\ \al\be}-\frac12g^{\mu\nu}
S^{\al\be\si}S_{\al\be\si}
\eeq
and where $^oG^{\mu\nu}$ is the Einstein tensor of Riemannian spacetime.
 This shows that, for example, a torsion wave carries energy and momentum. In fact, using the gauge invariance

\beq\label{gaugeinv}
\ps_{\mu\nu}\ra \ps_{\mu\nu}+\xi_{[\mu,\nu]}
\eeq

\no one may derive the wave equation from (\ref{tfe}), and it was shown that torsion waves carry energy.\cite{tp}

Since the torsion is totally antisymmetric we may equally well define the torsion vector by

\beq
b_\mu=\ep_{\mu\al\be\si}S^{\al\be\si}
.\eeq
Now, defining $b_\mu=\{b_0,\bm b\}$ the field equations for torsion may be written in the Minkowski space limit.\cite{hammondreview} They are written and compared to electromagnetism in the Table. The point to be made here is that the torsion obeys the wave equation. From the field equations in vacuum we have the constraint, 
${\bm k}\times {\bm b}=0$ so that
a traveling plane wave solution is given by

\beq\label{wave}
{\bm b}=b\sin(kz-\om t)\hat{\bm z}
.\eeq
It has been shown in \cite{tp} torsion waves are created by coherently flipping the spin of $N$ particles with spin at frequency $\om$ and that the power is proportional to $N^2\om^4$.

\begin{figure}

\begin{tabular}{|l|c|c|}\hline

& Torsion & Electromagnetism\\  \hline

Potential & ${\bm a}$, ${\bm A}$ & $\ph$, ${\bm A}$\\ \hline

Field &\raisebox{-.1em}{${\bm b} ={\bm \na }\times{\bm A }-\dot{\bm a }$}
&\raisebox{-.3em}{${\bm E }=-{\bm \na }\ph-\dot{\bm A }$}\\
 & $b_o={\bm \na}\cdot{\bm a}$ & ${\bm B}={\bm \na }\times{\bm A}$\\ \hline

Sources & ${\bm N}$, ${\bm I}$  &  ${\bm j}$, $\rh$\\ \hline

Field &  \raisebox{-.3em}{${\bm \na}b_o=-\dot{\bm b}-{\bm I}$}
&\raisebox{-.3em}{${\bm \na}\times{\bm H}=4\pi{\bm J}+\dot{\bm D}$}\\
Equations & ${\bm \na}\times{\bm b}={\bm N}$ &
 \raisebox{-.3em}{${\bm \na}\times{\bm E}+\dot{\bm B}=0$}\\
&&\raisebox{-.1em}{ ${\bm \na}\cdot{\bm D}=4\pi\rh$}\\
&& $ {\bm \na}\cdot{\bm B}=0 $\\ \hline

Gauge & $ {\bm a}\ra{\bm a}+{\bm \na}\times{\bm V}$ &\raisebox{-.2em} {$\ph\ra\ph-\dot\la$}\\
Invariance & ${\bm A}\ra{\bm A}+{\bm \na}\ph+\dot{\bm V}$ &
${\bm A}\ra{\bm A}+{\bm \na}\la$\\ \hline

Wave & \raisebox{-.3em}{$\na^2{\bm A}-\ddot{\bm A}=0$} &
\raisebox{-.3em}{ $\na^2{\bm A}-\ddot{\bm A}=0$} \\
Equation &&\\ \hline

\end{tabular}

\caption{\label{emtortab}Comparison of  torsion and electromagnetism.}

\end{figure}

\section{Dirac equation}

To describe spin one-half particles it is useful to introduce the tetrad
$e_\mu^{\ i}$ according to

\beq\label{geta}
g_{\mu\nu}=e_\mu^{\ i} e_\nu^{\ j}\eta_{ij}
\eeq
where
$\eta_{ij}$ is the Minkowski metric. Latin indices label the tetrad indices (there are four) and Greek are the coordinate indices. The tetrads are taken to be orthonormal so that
$e^\mu_{\ i} e_\mu^{\ j}=\de ^j_i.$
It is also useful to define
the object of anholonomity,

\beq
\Omega_{\al\be}^{\ \ a}=e_{[\be,\al]}^a
,\eeq
and

\beq
\Omega_{ab}^{\ \ c}\equiv e^\al_{\ a}e^\be_{\ b}\Omega_{\al\be}^{\ \ c}
.\eeq

The covariant derivative, in anholonomic coordinates,  is given by

\beq\label{ancovder}
\na_iA^j=A^j_{\ ,i}+\Ga_{im}^{\ \ \ j}A^m
\eeq
where  $\Ga_{im}^{\ \ \ j}$,
is the anholonomic connection coefficient,
and $\pa_i\equiv e_i^\mu\pa_\mu$. 

We may write $\na_iA^j
=e^\mu_{\ i}e_\nu^{\ j}\na_\mu A^\nu$ so that

\beq\label{affcon}
\Gamma_{\alpha\beta}^{\ \ \ \sigma}=\Gamma_{ab}^{\ \ m}
e_{m}^{\sigma}e_{\beta}^{b}e_{\alpha}^{a}-e_{b,a}^{\sigma}
e_{\beta}^{b}e_{\alpha}^{a}
.\eeq
This important result relates the affine connection $\Gamma_{\alpha\beta}^{\ \ \ \sigma}$
to the anholonomic connection
coefficients $\Gamma_{ab}^{\ \ m}$.  One may invert (\ref{affcon}) by transvexing
with the tetrad (and the dual tetrad),  which yields,

\beq\label{sc1}
\Ga_{ab}^{\ \ c}=e^\al_{\ a}e^\be_{\ b}e_\si^{\ c}\Ga_{\al\be}^{\ \ \si}
-e^\al_{\ b}e_{\al\ ,a}^{\ c}
.\eeq

Latin indices are raised and lowered with $\eta_{ab}$,  and we
define the quantity $\Ga_{\mu}^{\ ab}$:

\beq
\Ga_{\mu}^{\ ab}\equiv e_\mu^{\ n}\Ga_{n}^{\ ab}
.\eeq

\no Written as $\om_{\mu mn}=
\et_{am}\et_{bn}\Ga_{\mu}^{\ ab}$,
and $\om_{amn}=e_a^{\ \mu}\om_{\mu mn}$,
it is called the spin connection or the
connection 1-form.  We can also write
\beq
\Ga_{abc}=-\Omega_{abc}+\Omega_{bca}-\Omega_{cab}+S_{abc}
.\eeq

The curvature tensor is

\beq\label{anr}
R_{jki}^{\ \ \ n}=2(\Ga_{[j{\underline i},k]}^{\ \ n}
+\Ga_{[j{\underline i}}^{\ \ m}\Ga_{k]m}^{\ \ n}
+\Omega_{kj}^m\Ga_{mi}^{\ \ n})
\eeq

\no where brackets imply antisymmetrization with respect to the non-underlined indices.

With this we can write the covariant derivative of a spinor $\psi$ representing a spin one-half particle,

\beq\label{diracder}
D_\mu\psi=\psi_{,\mu}-\frac14\ga_a\ga_b\Ga_\mu^{\ ab}\psi
.\eeq
The  minimally coupled matter Lagrangian may now be written as

\beq\label{diraclag}
L=-\frac{i\hbar c}{2}\left[ (\overline{D_a\ps})\ga^a\ps-\overline\ps\ga^aD_a\ps
-\frac{2imc}{\hbar}{\overline\ps}\ps\right]
\eeq
with

\beq\label{adjdir}
\overline{D_a\ps}=\overline\psi_{,\mu}+\frac14\Ga_\mu^{\ ab}\overline\psi\ga_a\ga_b
.\eeq
with $e\equiv\sqrt{-g_{\mu\nu}}$.

We can now consider variations of the tetrad and the torsion potential,

\beq\label{vpd}
\de\int e\left(\frac{R}{ 2k}+L\right)d^4x=0
.\eeq
In (\ref{vpd}) variations are taken with respect to the
tetrad $e_{\mu}^i$, the Dirac adjoint $\overline\ps$, and
the torsion potential $\ps_{\mu\nu}$. The resulting
field equations are

\beq\label{gfetet}
G^{mn}-e^m_\mu e^n_\nu S^{\mu\nu\si}_{\ \ \ ; \si}
-2S^{mab}S^n_{\ ab}=-\frac{i\hbar ck}{ 2}\Si^{mn}
,\eeq

\beq\label{dirac2}
\ga^aD_a\psi+\frac{imc}{ \hbar}\psi=0
,\eeq
and

\beq\label{tfetet}
S^{\al\be\si}_{\ \ \ \ ;\si}=
-\frac{i\hbar ck}{2}\Xi^{\al\be\si}_{\ \ \ \ ;\si}
\eeq
where $\Xi^{\al\be\si}=(1/2)\ps\ga^{[\al}\ga^\be\ga^{\si]}\ps$.
In the above, the energy momentum tensor of the gravitational field
equations is defined according to

\beq\label{emdef}
\de\int e Ld^4x\equiv-\frac{i\hbar c}{ 2}\int e d^4x\Si_i^{\ \mu}\de e_\mu^i
,\eeq
which turns out to be
\begin{widetext}
\beq
\Xi_{mn}=\overline\psi\ga_{(m}D_{n)}\psi-D_{(m}\overline\psi\ga_{n)}
+\frac i3\overline\psi(\ga^ab_ag_{mn}-b_{(m}\ga_{n)})\psi
.\eeq
\end{widetext}
However, in the broadest context, one usually considers the non-minimally couple field in which case

\begin{widetext}
\beq\label{diraclag3}
L=-\frac{i\hbar c}{ 2}\left[ (D_a\overline\ps)\ga^a\ps-\overline\ps\ga^aD_a\ps
-\frac{2imc}{\hbar}\overline\ps\ps-\frac B4\overline\ps\gamma^5\gamma^\mu b_\mu\ps\right]
\eeq
\end{widetext}
where $B$ must be determined experimentally.

\section{Spin flip}
We are now in a position to consider the spin flip of an electron due to an incident torsion wave. We shall consider an electron trapped in a uniform magnetic field. 
It is assumed that pure gravitational effects are negligible and therefore we consider the torsion to be propagating in a Minkowski spacetime.

It is noted that this is as close as we can come to a free electron with spin. To account for spin there must be some fiduciary measurable, and the most reasonable case is that of a magnetic field. The solution to the Dirac equation for an electron trapped in a 
uniform field is given by\cite{huff}

\beq
\psi_f=C_fe^{-i(E_ft-p_x^fx-p_z^fz)}e^{-\xi^2/2}u_f
\eeq

\no where the momentum terms are eigenvalues, 

\beq
C_f^2=\frac{\sqrt{eB}(E_f+m)}{8L_xL_zE_f}
\eeq
and

\beq
\xi=\sqrt{eB}y -\frac{p_x}{\sqrt{eB}}
.\eeq
\no  We assume the final state is spin up and the initial state is spin down, so that,

\beq
u_f=\left(
\begin{array}{cccc}
h_{n-1} &\\
0 &\\
p_z^fh_{n-1}/(E_f+m) &\\
-\sqrt{2neB}h_n/(E_f+m) &\\
\end{array}\right)
\eeq
and

\beq
u_i=\left(
\begin{array}{cccc}
0 &\\
h_n &\\
-\sqrt{2neB}h_{n-1}/(E_i+m)  &\\
-p_z^ih_n/(E_i+m) &\\
\end{array}\right)
\eeq

\no where $h_n=N_nH_n$ where $H_n$ are the Hermite polynomials $N_n=1/\sqrt{2^nn!\sqrt{\pi}}$. By definition, here, the Hermite polynomial with a negative subscript is zero, and it is assumed that we have two dimensional box normalization, the sides being $L_x$ and $L_z$. The functions $u$ will be  functions of $y$ which goes from minus to plus infinity. It may be noted that the gauge freedom in the choice of the electromagnetic potential translate to a freedom in the choice of $y$ or $x$, or a combination of the two.

We start with the transition amplitude, defined by

\beq\label{ta}
S_{fi}=-i\frac{\kappa}{4}\int d^4 x\overline\psi_f\ga^5\ga^\si b_\si\psi_i
\eeq
where $\kappa=1+B$
and  $f$ and $i$ denote final and initial states and
$b_\si$ is the torsion field. Below, this will be compared to case when an electromagnetic waves interacts with the spin, in which case we have,

\beq
S_{fi}^\text {em}=-ie\int d^4 x\overline\psi_f\ga^\si A_\si\psi_i
.\eeq

To begin let us write (\ref{ta}) as

\beq\label{ta2}
S_{fi}=
-\frac{iC_fC_i}{L_0}\int d^4x 
e^{i(E_ft-p_x^fx-p_z^fz)}e^{-\xi^2}
M_{fi}e^{-i(E_it-p_x^ix-p_z^iz)}
\eeq
where
$L_0=\om/eE$ and
the matrix element is

\beq
M_{fi}=u_f^\dag\ga^0\ga^5\ga^\si b_\si u_i
,\eeq

which becomes

\begin{widetext}
\beqa\label{torm}
M_{fi}=b_0
   \left(\frac{h_{{n_f-1}}
   h_{n_i} \pi
   _z}{E_f+m}-\frac{h_{{n_f-1}} h_{n_i}
   p^f_z}{E_i+m}\right)
+b_1 \left(\frac{\sqrt{2}
   h_{{n_f-1}} h_{{n_i-1}}
   p^f_z \sqrt{B n_i
   e}}{(E_f+m)
   (E_i+m)}-\frac{\sqrt{2}
   h_{{n_f}} h_{n_i} \pi
   _z \sqrt{B n_f
   e}}{(E_f+m)
   (E_i+m)}\right)\\ \nonumber
   +b_2
   \left(\frac{i \sqrt{2}
 h_{{n_f-1}}  h_{{n_i-1}}
   p^f_z \sqrt{B n_i
   e}}{(E_f+m)
   (E_i+m)}+\frac{i \sqrt{2}
   h_{n_f} h_{n_i} \pi
   \sqrt{B n_f
   e}}{(E_f+m)
   (E_i+m)}\right)\\ \nonumber
   \\+b_3
   \left(-\frac{2 h_{n_f}
h_{{n_f-1}} \sqrt{B
   n_f e} \sqrt{B n_i
   e}}{(E_f+m)
   (E_i+m)}-\frac{h_{{n_f-1}} 
   h_{n_i} p^f_z
   \pi}{(E_f+m)
   (E_i+m)}+h_{{n_f-1}}
   h_{n_i}\right)   \nonumber.
   \eeqa  
\end{widetext}

It useful to compare this to the interaction of the electromagnetic field described by the potential $A_\si$, in which case (\ref{torm}) is replaced by

\begin{widetext}
\beqa\label{rich}
M_{fi}^\text {em}=A_0 \left(\frac{\sqrt{2}
   \sqrt{B (n+1) e} h_n h_{n+1}
   p_z^i}{(E_f+m)(E_i+m)}
   -\frac{\sqrt{2}
   \sqrt{B n e} h_{n-1} h_n
   p_z^f}{(E_f+m)
   (E_i+m)}\right)\\ \nonumber
+A_1
   \left(\frac{h_n^2
   p_z^f}{E_f+m}-\frac{h_n^2
   p_z^i}{E_i+m}\right)
+A_2 \left(\frac{i h_n^2
   p_z^i}{E_i+m}
   -\frac{i h_n^2
   p_z^f}{E_f+m}\right)\\ \nonumber
+A_3 \left(\frac{\sqrt{2} \sqrt{B
   (n+1) e} h_n
   h_{n+1}}{E_f+m}-\frac{\sqrt{2} \sqrt{B n e} h_{n-1}
   h_n}{E_i+m}\right)\\ \nonumber
.\eeqa   
\end{widetext}

In the case of electromagnetism we see the only non-zero matrix elements arise from the case in which the ``large" part of the spinor interacts with the``small" part, but with chiral torsion field there is the much larger term multiplied by $b_3$, which is not multiplied by $p/E$ (the last term in \ref{torm}). In others words, for the electromagnetic field the spin flip term enters as $v/c$, whereas with torsion it enters in the zeroth order.

Now, with the wave given by (\ref{wave}) the scattering amplitude may be written as

\beq\label{ta1}
S_{fi}=S_{fi}^++S_{fi}^-
\eeq
where

\beq\label{ta10}
S_{fi}^\pm=\pm\frac{\ka C_iC_fb}{8}\int d^4 x
e^{- it\de^\pm }e^{i\Delta p^xx}e^{i(\Delta p^z\pm k)z}
e^{-\xi^2}{\cal M}_{fi}
\eeq
and where $\de^\pm=E_i-E_f\pm\om$, $\Delta p^x=p^x_f-p^x_i$
and ${\cal M}_{fi}\overline\psi_f\ga^5\ga^3\psi_i$. Defining the delta functions produced by the $x$ and $z$ integrations as ${\cal L}_x$ and ${\cal L}_z^\pm$ we have
\beq
{\cal L}_x=\int dx e^{i\Delta p^xx}
\eeq
and
\beq
{\cal L}_z^\pm=\int dz e^{i\Delta (p^z \pm k)z}
\eeq
and
\beq
S_{fi}^\pm=\pm\frac{\ka C_iC_fb}{8}
{\cal L}_x{\cal L}_x^\pm\int dtdy
e^{- it\de^\pm }
e^{-\xi^2}h_n^2(\xi)
\eeq
where it was assumed the transition was described by $n\ra n+1$.
Finally,  squaring and integrating over the density of states $d\rho=(L_xL_z/(2\pi)^2)dp_xdp_z$, and taking the non-relativistic limit one obtains the transition probability $W$, i. e.,

\beq
W=\int|S_{fi}|^2d\rh
.\eeq
In the low velocity limit this becomes

\beq\label{w}
W=
\left(\frac{\ka b}{8}\right)^2
\left(
\frac{\sin\frac{\om-\om_0}{2}t}{\frac{\om-\om_0}{2c}}
\right)^2.
\eeq
This is result is very similar to the Rabi formula, which is\cite{rabi}

\beq
W=
\left(\frac{\mu E}{\hbar}\right)^2
\left(
\frac{\sin\frac{\om-\om_0}{2}t}{\frac{\om-\om_0}{2c}}
\right)^2.
\eeq
where $E$ is the electric field amplitude and $\mu_b=e\hbar/2mc$.

\section{summary}
The main result of this paper is given by (\ref{w}). It describes how an electron interacts with the torsion field and shows new kinds of measurements to detect torsion are possible. Many experiments over the years investigating  torsion\cite{ni} have bet set up experiments hoping to detect a force between spin polarized bodies. So far, all these have done is set an upper limit on torsion. One main limitation is that there is no way to increase the field strength once the polarization has reached saturation. In the present scheme, we may in principle increase the power of the torsion wave in a way that is only limited by the amount of equipment used. 
To detect the actual spin flip of an electron we may rely on electromagnetism. Considerations for the best possible arrangements are underway.

\ed